\documentclass[showpacs,preprintnumbers,amsmath,amssymb,onecolumn,12pt]{revtex4-1}
\usepackage{graphicx}
\usepackage{dcolumn}
\usepackage{bm}
\usepackage{subfigure}
\usepackage{float}
\usepackage{multirow}
\usepackage{booktabs}
\usepackage{amsmath}
\usepackage{cases}
\usepackage{latexsym,amssymb}
\usepackage[mathscr]{eucal}
\usepackage{lineno}
\usepackage{physics}
\usepackage[bookmarks=true,
   colorlinks=true,
   linkcolor=blue,
   urlcolor=blue,
   citecolor=blue,
   bookmarks=true,
   hyperindex=true
]{hyperref}
\providecommand{\U}[1]{\protect\rule{.1in}{.1in}}

\newcommand{\f}{\begin{equation}}
\newcommand{\ff}{\end{equation}}
\newcommand{\fa}{\begin{eqnarray}}
\newcommand{\ffa}{\end{eqnarray}}

\begin{document}
\allowdisplaybreaks[4]
\title{Spherical photon orbits around the Kerr-like black hole in Einstein-Bumblebee gravity}
\author{Shi-Yu Li$^{1}$}
\author{Song-Shan Luo$^{1}$}
\author{Zhong-Wen Feng$^{1}$}
\email{zwfengphy@163.com}
\affiliation{1. School of Physics and Astronomy, China West Normal University, Nanchong 637009, China}
\begin{abstract}
In this paper, we investigate the photon orbits around a Kerr-like black hole in Einstein-Bumblebee gravity, where Lorentz symmetry is spontaneously broken. By solving the Hamilton-Jacobi equation, we derive a sixth-order polynomial that governs the photon motion, explicitly dependent on the rotation parameter $u$, the Lorentz violation parameter $\ell$, and the effective inclination angle $v$. We analyze photon orbit configurations in polar, equatorial, and general inclined planes, identifying significant deviations from the Kerr solution. In the polar and equatorial planes, we identify distinct photon orbit configurations and analyze their dependence on model parameters. For general inclined orbits, we find a critical inclination angle $v$ that determines the number and location of photon orbits in both extremal and non-extremal cases. All photon orbits are radially unstable, and the critical impact parameter decreases with increasing Lorentz violation, potentially providing observable signatures to differentiate Einstein-Bumblebee gravity from general relativity.

\end{abstract}

\maketitle

\section{Introduction}
\label{intro}
Recent breakthroughs in observational astronomy have dramatically enhanced our capacity to explore the near-horizon structure of black holes. In 2019, the Event Horizon Telescope (EHT) collaboration released the first image of the shadow cast by the supermassive black hole at the center of galaxy M87, offering direct visual evidence for the existence of event horizons and strongly curved spacetimes~\cite{ref1,ref2,ref3}. In addition, the Laser Interferometer Gravitational Wave Observatory (LIGO) and its counterparts have detected numerous gravitational wave signals from binary black hole mergers, opening an unprecedented observational window into the strong-field regime of gravity~\cite{LIGOScientific:2016aoc}. In this process, the investigation of photon dynamics in the vicinity of black holes has become a key frontier linking theoretical predictions with observational features.

Photon orbits play a crucial role in characterizing the optical appearance of black hole, such as their shadows, ring-like features in accretion flows, and strong gravitational lensing patterns. In the spacetime of Schwarzschild black hole, the photon sphere is a critical surface where photons can orbit in unstable circular trajectories~\cite{Bardeen:1973xb,Chandrasekhar1985}. For rotating black holes, such as Kerr black hole, the frame dragging leads to nontrivial effects: photon trajectories become non-planar, non-circular, and exhibit significant asymmetries with respect to the rotation axis. In particular, equatorial and polar photon orbits behave very differently. Current research on photon orbits around rotating black hole mainly focused on the equatorial and polar planes, which is largely due to the mathematical simplifications afforded by symmetry, as well as the direct relevance of these configurations to astrophysical observations~\cite{bardeen1972rotating,bardeen1975lense,teo2021spherical}. However, realistic observations usually involve observers located at arbitrary angles relative to black holes~\cite{Hod2013,chatterjee2020observational,yang2012quasinormal}. Consequently, the investigations beyond the equatorial and polar planes are essential for a more comprehensive understanding of photon dynamics in the vicinity of black holes~\cite{lu2023ring,cui2023precessing,Carter1968}.

Current studies of black hole photon orbits generally adopt two schemes. The first and most widely used is numerical simulation, particularly the ray-tracing method, which numerically integrates the geodesic equations to track photon trajectories in complex spacetime backgrounds and produces observable images of black hole photon orbits. This scheme can incorporate complex accretion disk structures, magnetic fields, and varied observational angles, and has thus been extensively utilized~ \cite{Grenzebach:2014fha,Tsukamoto:2017fxq,Liu:2024lbi,Liu:2024lve,Zheng:2024ftk}. However, while numerical methods provide flexibility, they often lack clear insight into the intrinsic mathematical structure of photon orbits. Therefore, it is necessary to use analytical methods to obtain exact or semi-analytical forms of photon orbits by accurately solving the Hamilton-Jacobi equations. Such analytical methods fundamentally clarify the characteristics of the orbital structure, such as critical inclination, orbital bifurcation, orbital stability, and other key phenomena. Recent studies, notably those by Tavlayan and Tekin~\cite{Tavlayan2020,Chen2023,Alam:2024mmw,Tavlayan:2022hzl}, explicitly provided analytical solutions for spherical photon orbits in Kerr spacetimes, revealing richer and more intricate orbital structures, including critical inclination angles and bifurcation phenomena not readily captured by numerical simulations. Therefore, utilizing analytical schemes to accurately study photon orbits around black hole is important for elucidating their fine structure and revealing the geometric and physical mechanisms in the vicinity of strong gravitational fields.

However, both numerical and analytical methods typically assume classical general relativity (GR). Yet in high-energy regimes where quantum and gravitational effects become equally important, GR alone may be inadequate, and one must consider possible violations of Lorentz invariance (LI). One of the fundamental principles underlying our understanding of spacetime is LI, which serves as a cornerstone of both GR and the Standard Model (SM) of particle physics. GR offers an accurate description of gravitational phenomena on large scales, while SM effectively explains the behavior and interactions of elementary particles. Although these two frameworks are both well-established, they operate in different domains GR in the classical regime and the SM in the quantum regime,and their theoretical foundations and applicable energy ranges differ significantly. Nonetheless, growing theoretical and observational considerations suggest that LI may not be an exact symmetry at all energy scales~\cite{Mattingly2005}. In particular, the quest to unify gravity with quantum mechanics has led to the possibility that LI could be violated under extreme conditions. GR is constructed on a curved spacetime background and does not include quantum effects, whereas the SM is formulated on flat spacetime and neglects gravity. Because each theory omits key features of the other, they are fundamentally incompatible at very high energies, especially near the Planck scale. This incompatibility highlights the need for a unified framework that can consistently describe both gravitational and quantum phenomena.

According to predictions from GR, gravitational interactions between particles become increasingly significant above the Planck scale~\cite{AmelinoCamelia2013}, where neither GR nor SM alone suffices. Consequently, constructing a consistent framework that incorporates both quantum effects and gravity, namely a theory of quantum gravity, has become a central objective in theoretical physics. Moreover, LI is a continuous spacetime symmetry, which may not survive if spacetime exhibits a discrete structure at microscopic scales. This possibility motivates a re-evaluation of fundamental spacetime symmetries and has led to extensive research on Lorentz violation (LV). Recent studies on LV span a wide range of areas, including anomalies in neutrino propagation~\cite{Dai2017}, Lorentz-violating terms in the Standard model extension (SME)~\cite{Kostelecky1997, Kostelecky1998, Kostelecky2004}, LV effects in nongravitational sectors~\cite{Coleman1997, Coleman1999, Myers2003}.

Among them, the SME is an effective field theory framework designed to describe the coupling between the SM and GR. It accommodates dynamical spacetime curvature and includes additional terms that encode possible LV effects originating at the Planck scale~\cite{Kostelecky1997}. In the SME, the LV terms typically take the form of Lorentz violating parameter coupled to background coefficients carrying Lorentz indices. The appearance of a nonzero vacuum expectation value  for one or more quantities with local Lorentz indices is considered a signal of local Lorentz symmetry breaking. A concrete example is the so called bumblebee model, in which LV arises from the dynamics of a single vector or axial-vector field $B_\mu$, known as the bumblebee field. This model represents a special case of Einstein-aether theory, with dynamics governed by a potential that attains a non-zero minimum, thereby inducing a spontaneous Lorentz symmetry breaking via the vacuum expectation value of the field. The bumblebee gravity model was first proposed by Kosteleck and Samuel in 1989~\cite{Kostelecky1989a, Kostelecky1989b} as a simple framework for spontaneous LV and has since attracted considerable interest in the theoretical physics community~\cite{Ovgun2019, Casana2018, Capelo2015, Jesus2019, Santos2015, Hernaski2014, Maluf2015, Ding2019}, Following significant progress in the study of Lorentz-violating theories, exact solutions describing Kerr-like black holes have been derived within the framework of Einstein-Bumblebee gravity ~\cite{ Ovgun2018, Oliveira2021, Kanzi2019, Gullu2022, Maluf2021, Ding2022, Ovgun2019,islam2024investigating}.

It is worth emphasizing that although Kerr-like black hole in Einstein-Bumblebee gravity have attracted increasing attention, most existing studies on their photon orbits rely primarily on numerical simulations\cite{Ding2019,Liu:2019mls,Wang:2021irh}, making it difficult to capture the subtle effects of the LV parameter and the rotation parameter on the orbital structure. To provide a clearer and more precise understanding of these physical effects, this work systematically analyzes spherical photon orbits in Kerr-like black hole in the Einstein–Bumblebee gravity framework, including polar, equatorial, and general  orbits. In particular, special attention is given to the role of the LV parameter $\ell$. Starting from the gravitational sector, we derive analytical expressions for the photon orbits, offering more accurate and transparent insights than purely numerical approaches. Our results highlight the deviations from the standard Kerr scenario, emphasizing the influence of Lorentz symmetry breaking.

The structure of this paper is organized as follows. Section~\ref{sec2} briefly reviews the Kerr-like black hole solution in Einstein-bumblebee gravity, and derive the governing equation for spherical photon orbits from the Hamilton-Jacobi formalism, resulting in a sixth-order polynomial equation that involves the orbit radius, the rotation parameter $u$, the Lorentz violating parameter $\ell$, and the effective inclination angle $v$. Section~\ref{sec3} investigates photon trajectories confined to the polar and equatorial planes. In Section~\ref{sec4}, we extend the analysis to general spherical photon orbits located between these planes. Finally, Section~\ref{sec5} summarizes the main results and provides concluding remarks.

\section{The Kerr-like black hole solution in Einstein-bumblebee gravity and its equations of motion}
\label{sec2}

In the theory of Einstein-Bumblebee, the vector field $B_\mu$ acquires a non-zero vacuum expectation value through a suitable potential, thereby inducing spontaneous Lorentz symmetry breaking in the gravitational sector~\cite{Bluhm2005,Kostelecky1989a}. The action is given by
\begin{equation}
\label{eq1}
S = \int {\rm{d}}^4x \sqrt{-g} \left[ \frac{1}{16\pi G_N} \left( R + \varrho B^\mu B^\nu R_{\mu\nu} \right) - \frac{1}{4} B_{\mu\nu} B^{\mu\nu} - V \left(B^\mu B_\mu \pm b^2 \right) \right],
\end{equation}
where $\varrho$ is a real coupling constant controlling the non-minimal interaction between gravity and the Bumblebee field $B_\mu$, which has mass dimension one. The potential $V$ is minimized when $B^\mu B_\mu \pm b^2 = 0$ with $b^2 > 0$. The field strength tensor is defined as $B_{\mu\nu} = \partial_\mu B_\nu - \partial_\nu B_\mu$. According to the action~(\ref{eq1}), the metric of a Kerr-like black hole in Einstein-Bumblebee gravity, expressed in standard Boyer-Lindquist coordinates, is given by
\begin{align}
\label{eq8}
{\rm{d}}s^2 & =  -\left(1 - \frac{2Mr}{\rho^2} \right) {\rm{d}}t^2 + \frac{\rho^2}{\Delta} {\rm{d}}r^2 + \rho^2 {\rm{d}}\theta^2 + \frac{{\cal A} \sin^2 \theta}{\rho^2} {\rm{d}}\phi^2
\nonumber \\
& - \frac{4 M r a \sqrt{1 + \ell} \sin^2 \theta}{\rho^2} {\rm{d}}t {\rm{d}}\phi,
\end{align}
where
\begin{subequations}
\begin{align}
\label{eq9}
\rho^2 &= r^2 + \left(1 + \ell \right) a^2 \cos^2 \theta, \\
\label{eq10}
\Delta &= \frac{{{r^2} - 2Mr}}{{1 + \ell }} + a^2, \\
\label{eq11}
{\cal A} &= \left[r^2 + \left(1 + \ell \right) a^2 \right]^2 - \Delta \left(1 + \ell \right)^2 a^2 \sin^2 \theta,
\end{align}
\end{subequations}
and $a$ is the rotation parameter of the black hole, $\ell$ is the LV parameter arising from the vacuum expectation value of the Bumblebee field, modifying the standard Kerr metric structure accordingly. For $\ell=0$, Eq.~(\ref{eq8}) recovers the classical Kerr metric.

We now consider the equations of motion for photons around a black hole, which follow null geodesics. These equations can be conveniently derived using the Hamilton–Jacobi method, which allows separation of variables in axisymmetric spacetime. In the Einstein–Bumblebee gravity background, the Hamilton–Jacobi equation takes the form
\begin{equation}
\frac{\partial S}{\partial\lambda}=-\frac12g^{\mu\nu}\frac{\partial S}{\partial x^\mu}\frac{\partial S}{\partial x^\nu},
\label{eqJacobi}
\end{equation}
where $\lambda$ denotes the affine parameter along the null geodesics. Since the spacetime of the black hole  is axisymmetric, it possesses the Killing vectors ${\xi ^t}$ and ${\xi ^\theta}$, which are related to the photon energy $E$ and the $z$-component of the photon angular momentum $L_z$, respectively. According to the two Killing vectors, the Jacobi action $S$ can be expressed as follows
\begin{equation}
S = -E t + L_z \phi + S_r \left(r\right) + S_\theta \left(\theta \right),
 \label{eq:action_s}
\end{equation}
where $S_r\left(r\right)$ and $S_\theta\left(\theta\right)$ are the separated parts of the Jacobi action, which depend only on the radial coordinate $r$ and the angular coordinate $\theta$, respectively.  By combining Eq.~(\ref{eqJacobi}) with Eq.~(\ref{eq:action_s}), one has

\begin{subequations}
\label{eq16}
\begin{align}
\left(1 + \ell \right) \rho^2 \frac{{\rm{d}}t}{\rm{d}\lambda} &=
a L_z \left(1 + \ell \right) - a^2 \left(1 + \ell \right)^{\frac{3}{2}} E \sin^2 \theta  \nonumber\\
&  + \frac{r^2 + \left(1 + \ell \right) a^2}{\Delta} \left\{ \left[r^2 + \left(1 + \ell \right) a^2 \right] E - a L_z \sqrt{1 + \ell} \right\},  \\
\left( {1 + \ell } \right){\rho ^2}\frac{{{\rm{d}}\phi }}{{{\rm{d}}\lambda }} &= \frac{{\left( {1 + \ell } \right){L_z}}}{{{{\sin }^2}\theta }} - a{\left( {1 + \ell } \right)^{\frac{3}{2}}}E \nonumber\\
& + \frac{{a\sqrt {1 + \ell } }}{\Delta }\left\{ {\left[ {{r^2} + \left( {1 + \ell } \right){a^2}} \right]E - a{L_z}\sqrt {1 + \ell } } \right\},\\
\rho^2 \frac{{\rm{d}}r}{{\rm{d}}\lambda} &= \sqrt{\mathcal{R}\left(r \right)}, \\
\rho^2 \frac{\rm{d}\theta}{{\rm{d}}\lambda} &= \sqrt{\Theta \left(\theta \right)}.
\end{align}
\end{subequations}
where the radial $\mathcal{R} \left(r \right)$ and angular motion $\Theta \left(\theta \right)$ of photons are given by
\begin{subequations}
\label{eq17}
\begin{align}
\mathcal{R}\left( r \right) \! & =\! - \Delta \left[ {{{\cal K}} + L_z^2 + {a^2}{E^2}\left( {1 + \ell } \right) - 2a{L_z}E\sqrt {1 + \ell } } \right] \!+\!{\left\{ {\frac{{\left[ {{r^2} + {a^2}\left( {1 + \ell } \right)} \right]E}}{{\sqrt {1 + \ell } }} - a{L_z}} \right\}^2}, \label{eq17a}\\
\Theta \left( \theta  \right) & = {{\cal K}} + \left( {1 + \ell } \right){a^2}{E^2}{\cos ^2}\theta  - L_z^2{\cot ^2}\theta ,\label{eq17b}
\end{align}
\end{subequations}
with the Carter constant $\mathcal{K}$ that arises due to the separability of the geodesic equations in axisymmetric spacetime. In the radial direction, unstable photon orbits are characterized by the conditions $\mathcal{R} \left(r\right) = 0$ and ${{{\rm{d}}\mathcal{R}\left( r \right)} \mathord{\left/ {\vphantom {{{\rm{d}}\mathcal{R}\left( r \right)} {{\rm{d}}r}}} \right. \kern-\nulldelimiterspace} {{\rm{d}}r}} = 0$. By simultaneously solving these conditions together with Eq.~(\ref{eq17a}) , one obtains two critical conserved parameters are obtained as follows~\cite{Carter1968,Ding2019}:
\begin{subequations}
\label{eq19}
\begin{align}
\xi \left( r \right) & = \frac{{{L_z}}}{E} = \frac{{{r^2}\left( {3M - r} \right) - \left( {1 + \ell } \right){a^2}\left( {M + r} \right)}}{{\sqrt {1 + \ell } {\mkern 1mu} a\left( {r - M} \right)}}, \\
\eta \left( r \right) & = \frac{{{\cal K}}}{{{E^2}}} = \frac{{{r^3}\left[ {4\left( {1 + \ell } \right){\mkern 1mu} M{a^2} - r{{\left( {r - 3M} \right)}^2}} \right]}}{{\left( {1 + \ell } \right){\mkern 1mu} {a^2}{{\left( {r - M} \right)}^2}}}.
\end{align}
\end{subequations}
In order to characterize the deviation between the photon orbital plane and the equatorial plane of the black hole, it is necessary to introduce the effective inclination angle, which can be defined as follows~\cite{Hod2013}
\begin{equation}
\cos i = \frac{L_z}{\sqrt{L_z^2 + \mathcal{K}}},
\label{eq20}
\end{equation}
where angle $i$ effectively measures the tilt of the orbital plane relative to the equatorial plane of the black hole. When the photon orbital plane coincides exactly with the equatorial plane of the black hole, the inclination angle vanishes, which implies $\cos i = 1$ and $\mathcal{K} = 0$. In contrast, when the photon travels along a polar trajectory perpendicular to the equatorial plane, its azimuthal angular momentum vanishes ($L_z = 0$), and thus $\cos i = 0$.

To facilitate further analysis, we non-dimensionalize the key variables by scaling with respect to the mass $M$ of the black hole. We define the following dimensionless parameters:
\begin{equation}
x = \frac{r }{M}, \quad u = \frac{a^2}{M^2}, \quad v = \sin^2 i.
\label{eq21}
\end{equation}
In the context of this work, we refer to $u$ and $v$ as the rotation parameter and the effective inclination angle parameter, respectively. The physically reasonable ranges for the relevant variables are given as follows
\begin{equation}
x \geq 0, \quad 0 \leq u \leq 1, \quad 0 \leq v \leq 1, \quad \ell > -1.
\label{eq22}
\end{equation}

By combining the effective potential method with the equations of motion in Kerr-like geometries, the radial behavior of photons can be derived as a function of the rotation and inclination parameters. By incorporating Eqs.~\eqref{eq19} and \eqref{eq20}, the photon motion condition can be expressed as the following a sextic
equation~\cite{Hod2013,Tavlayan2020}
\begin{align}
f\left( x \right) & = {\left( {1 + \ell } \right)^2}{u^2}v + 2{\left( {1 + \ell } \right)^2}{u^2}vx + \left( {1 + \ell } \right)u\left[ {\left( {1 + \ell } \right)u - 6} \right]v{x^2} \nonumber \\
& - 4 \left(1 + \ell \right) u x^3 + \left[9 + 2 \left(1 + \ell \right) u v\right] x^4 - 6x^5 + x^6.
\label{eq23}
\end{align}
It is evident that the roots of the polynomial~(\ref{eq23}) correspond to the photon orbits in the black hole spacetime. However, obtaining analytical solutions for these roots is generally intractable. Therefore, in the following sections, we will examine the properties of photon orbits by considering three representative cases.

\section{Analysis of photon orbits in equatorial and polar planes}
\label{sec3}
\subsection{The photon orbits on the polar plane}
\label{sec3-1}
On the polar orbital plane, that is $i =  \pm {\pi  \mathord{\left/ {\vphantom {\pi  2}} \right. \kern-\nulldelimiterspace} 2}$ or $ v = 1$, one as $ L_z = 0 $, leading to Eq.~(\ref{eq23}) reduces to a cubic polynomial as follows
\begin{equation}
\label{eq24}
f \left(x\right)=x^3 -3x^2 +\left(1+\ell \right)ux +\left(1+\ell \right)u,
\end{equation}
which give three real roots $x_1$, $x_2$, and $x_3$. Furthermore, the event horizon $x_h$ of the Kerr-like black hole in Einstein-Bumblebee gravity can be expressed as
\begin{equation}
x_h = 1 + \sqrt{1 - (1 + \ell) u}.
\label{eq25}
\end{equation}

\begin{figure}[htbp]
\centering
\includegraphics[width=0.65\textwidth]{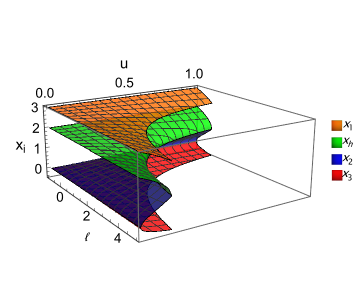}
\vspace{-1.5\baselineskip}
\caption{The polar photon orbits $x_i$ and event horizon $x_h$ of the Kerr-like black hole in Einstein-Bumblebee gravity as the functions of the rotation parameter $u$ and LV parameter $\ell$.}
\label{fig1}
\end{figure}
According to Eqs.~(\ref{eq24}) and (\ref{eq25}), all the polar photon orbits, as well as the event horizon (green surface) of the Kerr-like black hole in Einstein-bumblebee gravity, are plotted as functions of the rotation parameter $u$ and the LV parameter $\ell$ in Fig.~\ref{fig1}. As illustrated, one of the photon orbits, denoted by $x_1$, exists outside the event horizon $x_h$, while the other two photon orbits, $x_2$ and $x_3$, are located inside the horizon. It is important to note that $x_3$ takes negative value and is thus unphysical. Consequently, such unphysical root is excluded from the subsequent analysis. Although only photon orbits located outside the event horizon are likely to contribute to observable phenomena. However, the inner photon sphere also has physical significance because it can reflect some fundamental properties of a black hole. Therefore, from the perspective of theoretical research, it is also worthwhile to consider the photon orbits located within the event horizon.

\begin{figure}[htbp]
\centering
\includegraphics[width=0.8\textwidth]{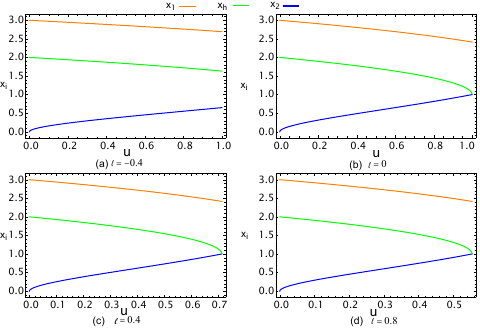}
\caption{The behavior of polar photon orbits ($x_1$ and $x_2$) and event horizon $x_h$ versus the rotation parameter $u$ for different LV parameter $\ell$.}
\label{fig2}
\end{figure}

To further investigate the influence of different parameters on the structure and dynamics of photon orbits of the Kerr-like black hole in Einstein-Bumblebee gravity, we present Fig.~\ref{fig2}, which provides a more detailed parametric analysis. The overall trend shows that, for a fixed $\ell$, the radius of the outer photon orbit $x_1$ and the event horizon $x_h$ decrease monotonically with increasing $u$, while the radius of the inner orbit $x_2$ increases monotonically. However, when $\ell < 0$,  as shown in Fig.~\ref{fig2}(a), for $\ell = -0.4$, and in the extreme case of $u = 1$, the event horizon $x_h$ and the inner photon orbit $x_2$ do not merge. In Fig.~\ref{fig2}(b), the case $\ell = 0$ corresponds to the classical Kerr black hole. As the LV parameter $\ell$ increases, both the photon orbits and the event horizon gradually deviate from the classical case. Moreover, as $\ell$ increases, the allowed range of the rotation parameter $u$ shrinks significantly. In the extremal case, i.e., $u = {1 \mathord{\left/
 {\vphantom {1 {\left( {1 + \ell } \right)}}} \right.
 \kern-\nulldelimiterspace} {\left( {1 + \ell } \right)}}$, the photon orbit that originally lies inside the event horizon coincides with the event horizon radius. Therefore, for an extremal Kerr-like black hole in Einstein-bumblebee gravity, the photon orbit within the horizon lies exactly on the horizon itself.

\begin{figure}[htbp]
\centering
\includegraphics[width=0.7\linewidth]{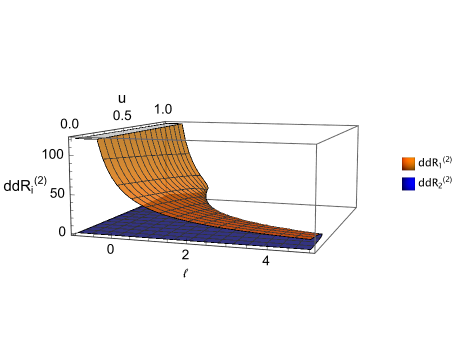}
\vspace{-1.5\baselineskip}
\caption{The ${\text{ddR}}_i^{\left( {2} \right)}$ as functions of $\ell$ and $u$ for the photon orbits $x_i$ on the polar plane.}
\label{fig3}
\end{figure}

Next, we analyze the stability of photon orbits on the polar plane, which determines whether photons are repelled from the black hole or fall into it under radial perturbations. According to Ref.~\cite{Cunha2017}, the stability of photon orbits is determined by
\begin{equation}
{\text{ddR}}_i^{\left( {2} \right)} = {\left. {\frac{{{\text{d}^2}R\left( x \right)}}{{\text{d}{x^2}}}} \right|_{x = {x_i}}},
\label{eq30+}
\end{equation}
where $R\left( x \right) = {{R\left( r \right)} \mathord{\left/ {\vphantom {{R\left( r \right)} {{M^4}{E^2}}}} \right. \kern-\nulldelimiterspace} {{M^4}{E^2}}}$. It is well known that ${\text{ddR}}_i^{\left( {2} \right)} < 0$, indicates a stable photon orbit, whereas ${\text{ddR}}_i^{\left( {2} \right)} > 0$ represents an unstable photon orbit, ${\text{ddR}}_i^{(2)} = 0$ corresponds to a saddle point. By substituting the roots of Eq.~(\ref{eq20}) into Eq.~(\ref{eq30+}), we depict the behavior of ${\text{ddR}}_i^{\left( {2} \right)}$ as functions of $u$ and $\ell$ in Fig.~\ref{fig3}. The orange and blue surfaces illustrate the stability of the photon orbits $x_1$ and $x_2$, respectively. It is observed that ${\text{ddR}}_1^{(2)}$ remains positive, indicating that the polar photon orbit $x_1$ is unstable under radial perturbations, while ${\text{ddR}}_2^{(2)}$ remains zero, implying that this root corresponds to a saddle point. Since the photon orbit $x_2$ resides within the event horizon, it cannot be observed by detectors at infinity. In contrast, the photon orbit $x_1$ lies outside the event horizon and is observable. Therefore, it is essential to analyze the critical impact parameter of $x_1$.

The impact parameter is a key physical quantity that describes the motion of particles or light within the gravitational field of black hole.	For rotating black hole, the inclusion of rotation and LV parameters introduces asymmetry and complexity to photon trajectories. Consequently, it is crucial to study the behavior of light photons originating from infinity as they traverse the vicinity of a Kerr-like black hole in Einstein-bumblebee gravity.	As described in Refs.~\cite{Carter1968,Tavlayan2020}, the critical impact parameter $\beta$ for a four dimensional axisymmetric spacetime is defined as
\begin{equation}
\beta = \frac{L}{ME} = \sqrt{\xi^2 + \eta},
\label{eqa31+}
\end{equation}
where $L$ is the angular momentum of photons. Substituting Eq.~(\ref{eq19}) into Eq.~(\ref{eqa31+}), the critical impact parameter of the Kerr-like black hole in Einstein-Bumblebee gravity is given by
\begin{equation}
\beta  = \sqrt {\frac{{2\left[ {{{\left( {1 + \ell } \right)}^2}{u^2}\left( {{x^2} + 1} \right) + {x^4}\left( {{x^2} - 4x + 5} \right) + 2\left( {1 + \ell } \right)u{x^2}\left( {{x^2} - 2x - 1} \right)} \right]}}{{\left( {1 + \ell } \right)u{{\left( {x - 1} \right)}^2}}}} .
\label{eq36+}
\end{equation}

\begin{figure}[htbp]
\centering
\vspace{-1.5\baselineskip}
\includegraphics[width=0.6\linewidth]{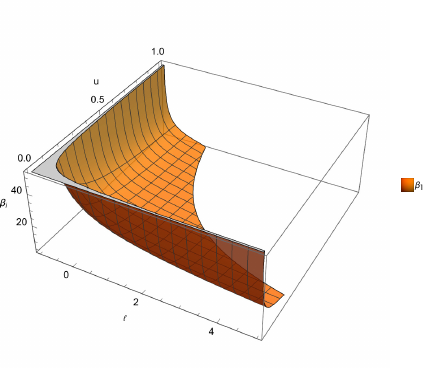}
\vspace{-1\baselineskip}
\caption{The critical impact parameter $\beta_i$ for polar orbits  as functions of $\ell$ and $u$.}
\label{fig4}
\end{figure}

Fig.~\ref{fig4} illustrates the critical impact parameter $\beta_1$ for photons on the polar orbit $x_1$ in a Kerr-like black hole in Einstein-bumblebee gravity, as derived from Eq.~(\ref{eq36+}). This parameter determines the possible trajectories of photons near the black hole: (i) Photons with impact parameters smaller than the critical value $\beta_1$ are captured by the black hole; (ii) Photons with impact parameters equal to $\beta_1$ enter the unstable orbit $x_1$, orbiting the black hole several times before eventually escaping;  (iii) Photons with impact parameters greater than $\beta_1$ are scattered to infinity and can be detected by distant observers.  In addition, the surface $\beta_1$ decreases with increasing values of $\ell$, indicating that the Einstein-bumblebee modification significantly weakens the black hole's ability to capture photons. In other words, more photons are scattered outward, enhancing their chances of detection. As a result, the polar photon ring of a Kerr-like black hole in Einstein-bumblebee gravity is expected to appear brighter than that of a classical Kerr black hole, which could provide observable signatures for testing deviations from GR.

\subsection{The photon orbits on equatorial plane}
\label{sec3-2}
\begin{figure}[htbp]
\centering
\includegraphics[width=0.6\linewidth]{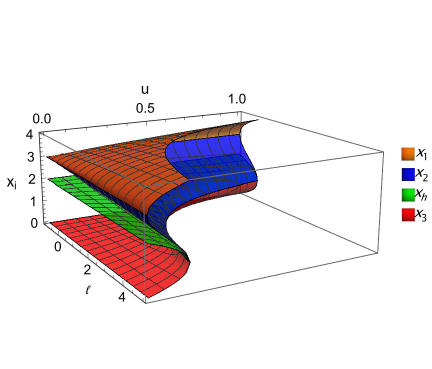}
\vspace{-1.5\baselineskip}
\caption{The equatorial photon orbits $x_i$ and event horizon $x_h$ of  Kerr-like black hole in Einstein-bumblebee gravity as the functions of the rotation parameter $u$ and LV parameter $\ell$.}
\label{fig5}
\end{figure}

For photon orbits of the equatorial plane, one has $ i = 0 $ or $\pi $, which leads to $ v = 0$ and $ \mathcal{K} = 0 $. Therefore, Eq.~(\ref{eq23}) is simplified as follows
\begin{equation}
f\left(x\right) =  x^3 - 6x^2 + 9x -4 \left(1 + \ell \right) u.
\label{eq31}
\end{equation}
By solving Eq.~(\ref{eq31}), one can obtain four different real roots in the non-extremal case, denoted in descending order as $x_1$, $x_2$, and $x_3$, respectively. Based on these roots, Fig.~\ref{fig5} illustrates the behavior of equatorial photon orbits for various values of the parameters $\ell$ and $u$.

As shown in Fig.~\ref{fig5}, when $\ell = 0$, the scenario reduces to the classical Kerr black hole case, where only three equatorial photon orbits exist in the spacetime: the retrograde orbit $x_1$, the prograde orbit $x_2$, and the inner orbit $x_3$. The retrograde orbit $x_1$ and the inner orbit $x_3$ increase monotonically with the rotation parameter $u$, while the prograde orbit $x_2$ and the event horizon radius $x_h$ decrease monotonically~\cite{teo2021spherical}. The introduction of the LV parameter $\ell$ significantly alters the characteristics of all equatorial photon orbits. As the parameter $\ell$ increases, the retrograde orbit $x_1$ and the inner orbit $x_3$ gradually increase, while the prograde orbit $x_2$ and the radius of the event horizon $x_h$ gradually decrease.

\begin{figure}[htbp]
\centering
\includegraphics[width=0.8 \linewidth]{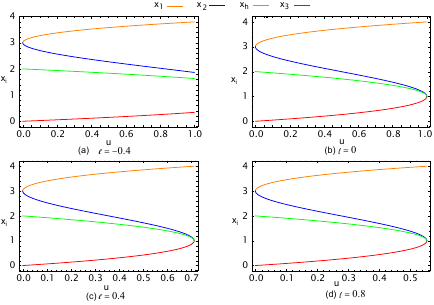}
\caption{The behavior of equatorial photon orbits $x_i$ and event horizon $x_h$ versus the rotation parameter $u$ for different LV parameter $\ell$.}
\label{fig6}
\end{figure}

This behavior arises primarily because increasing $\ell$ reduces the allowed upper bound of the rotation parameter $u$ (see Fig.~\ref{fig6}), thereby decreasing the angular momentum lost by photons in retrograde motion and gained in prograde motion. Moreover, in the extremal case (i.e., $u = {1 \mathord{\left/ {\vphantom {1 {\left( {1 + \ell } \right)}}} \right. \kern-\nulldelimiterspace} {\left( {1 + \ell } \right)}}$), the photon orbits $x_2$, $x_3$, and the radius of the event horizon $x_h$ coincide. Hence, the extremal Kerr-like black hole in Einstein-Bumblebee gravity possess only two equatorial photon orbits: one $x_1$ located outside the event horizon and the other $x_2$ lying exactly on the event horizon.

\begin{figure}[htbp]
\centering
\includegraphics[width=0.65\linewidth]{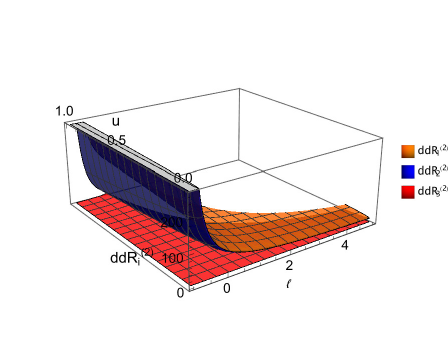}
\vspace{-1.5\baselineskip}
\caption{The ${\text{ddR}}_i^{\left( {2} \right)}$ as functions of $\ell$ and $u$ for equatorial photon orbits $x_i$.}
\label{fig7}
\end{figure}

Next, we examine the radial stability of equatorial photon orbits using Eq.~(\ref{eq30+}).	As shown in Fig.~\ref{fig7}, it can be seen that ${\text{ddR}}_1^{(2)} > 0$ and ${\text{ddR}}_2^{(2)} > 0$, while ${\text{ddR}}_3^{(2)} = 0$, indicating that the photon orbits $x_1$ and $x_2$ outside the event horizon are radially unstable, whereas the photon orbit $x_3$ inside the event horizon corresponds to a saddle point.

\begin{figure}[htbp]
\centering
\includegraphics[width=0.65\linewidth]{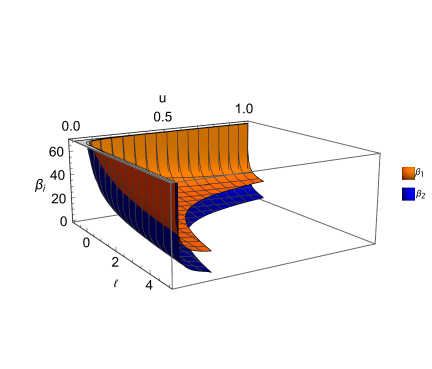}
\vspace{-1\baselineskip}
\caption{The critical impact parameters $\beta_i$  for photons in the retrograde and prograde orbits on the equatorial plane of the Kerr-like black hole in Einstein-Bumblebee gravity.}
\label{fig8}
\end{figure}

To analyze the behavior of photons near the retrograde and prograde orbits, we calculate the critical impact parameters of the photon orbits using Eq.~(\ref{eq36}) and present the results in Fig.~\ref{fig8}. The orange and blue surfaces correspond to the critical impact parameters $\beta_i$ of photons on the retrograde and prograde orbits, respectively, on the equatorial plane of the Kerr-like black hole in Einstein-bumblebee gravity. It is observed that as the rotation parameter $u$ increases, both the critical impact parameters $\beta_1$ for the retrograde orbit and $\beta_2$ for the prograde orbit decrease, with $\beta_1$ always remaining greater than $\beta_2$. This indicates that, in the background of a rotating black hole, photons in the prograde orbit are more likely to be scattered to distant detectors than those in the retrograde orbit, making the photon ring of the prograde orbit appear brighter. Moreover, the LV parameter $\ell$ restricts the allowed range of the rotation parameter $u$ and overall reduces the critical impact parameters for both the prograde and retrograde orbits. Consequently, compared to the classical Kerr black hole, the photon rings in both prograde and retrograde orbits on the equatorial plane of the Kerr-like black hole in Einstein-bumblebee gravity appear brighter.

\section{General photon orbits of the Kerr-like black hole in Einstein-Bumblebee gravity}
\label{sec4}
\begin{figure}[htbp]
\centering
\includegraphics[width=0.6\linewidth]{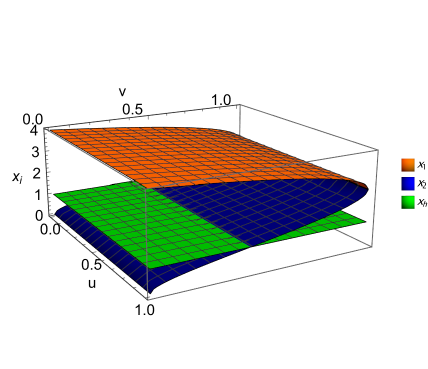}
\vspace{-1.8\baselineskip}
\caption{The general photon general orbits ($x_1$ and $x_2$) and event horizon $x_h$ of extreme Kerr-like black hole in Einstein-Bumblebee gravity as functions of the parameters $v$ and $u$.}
\label{fig9}
\end{figure}

In this section, we analyze the most general scenario of spherical photon orbits ($0<v<1$) around the Kerr-like black hole in Einstein-Bumblebee gravity. First, we examine the extreme case, where $u = {1 \mathord{\left/
 {\vphantom {1 {\left( {1 + \ell } \right)}}} \right.
 \kern-\nulldelimiterspace} {\left( {1 + \ell } \right)}}$ with $0 < u < 1$. Substituting $\ell$ in Eq.~(\ref{eq23}) with $u$, one obtains
\begin{align}
\label{eq35}
{\left( {x - 1} \right)^2}{\mathcal{P}_4}\left(x\right) = 0,
\end{align}
where
\begin{align}
{\mathcal{P}_4}\left( x \right) = \left( {x - 4} \right){x^3} + v\left( {2{x^2} + 4x + 1} \right).
\label{eq36}
\end{align}
It is observed that the double root $(x - 1)^2$ in Eq.~(\ref{eq35}) corresponds to the radius of the event horizon, which takes the value $x = 1$. According to Descartes' rule of signs, the quartic equation ${\mathcal{P}_4}\left( x \right) = 0$ can have at most two positive real roots, denoted as $x_1$ and $x_2$. Since the analytical expressions of these roots are relatively complicated, their dependence on the parameters $v$ and $u$ is illustrated in Fig.~\ref{fig9}.

As shown in Fig.~\ref{fig9}, there exists a critical inclination angle parameter $v_{\text{cr}}$ that governs the distribution of photon orbits in the context of Kerr-like black hole in Einstein-Bumblebee gravity. When the parameter satisfies $v  > {3 \mathord{\left/
 {\vphantom {3 7}} \right. \kern-\nulldelimiterspace} 7}$ (or inclination angle satisfies $\cos i < \sqrt {{4 \mathord{\left/ {\vphantom {4 7}} \right. \kern-\nulldelimiterspace} 7}} $), two photon orbits always exist outside the event horizon $x_h$, regardless of the value of the rotation parameter $u$, the prograde photon orbit $x_1$ and the retrograde photon orbit $x_2$. In contrast, when the inclination angle parameter is smaller, $v < v_{\text{cr}}$, only the retrograde photon orbit $x_2$ remains outside the event horizon. The prograde orbit $x_1$, in this case, either lies within the event horizon or coincides with it, and thus no longer exists as a distinct stable orbit. This reveals that variations in the inclination angle can significantly influence the structure and number of photon orbits, highlighting the geometric features of photon motion under Einstein-bumblebee gravity modifications.

Next, we focus on the non-extremal case. It is well known that algebraic equations of degree five or higher generally do not admit analytic solutions. Therefore, the sextic polynomial~(\ref{eq23}) cannot be solved analytically. Although numerical methods can be used to determine the radius of the photon orbit, these methods yield solutions in the form $x = \left(u, \ell, v\right)$ rather than the desired function $x = \left( u, \ell \right)$ that depends only on $u$ and $\ell$. According to Ref.~\cite{Hod2013}, there exists a critical inclination angle ${v_{\rm{cr}}} = v_{\rm{cr}}\left(u, \ell \right)$ in the parameter space $\left(u, \ell, v\right)$. When the inclination angle parameter is below this critical value, the equation has four real roots. When it is above, the number of real roots reduces to two. To determine this critical inclination angle, the parameter $v$ must be expressed as a function of $u$ and $\ell$ as follows
\begin{align}
v = \frac{\Xi }{u}.
\label{eq33}
\end{align}
At the critical point, one has
\begin{align}
\label{eq33-2}
{\Xi _{\rm{cr}}} = - \frac{{3{{\left( {w - 1} \right)}^3}}}{{\left( {1 + \ell } \right)\left[ {7 + (w - 5)w} \right]}},
\end{align}
with
\begin{align}
w = {\left[ {1 - \left( {1 + \ell } \right)u} \right]^{1/3}}.
\label{eq33-3}
\end{align}

\begin{figure}[htbp]
\centering
\includegraphics[width=0.6\linewidth]{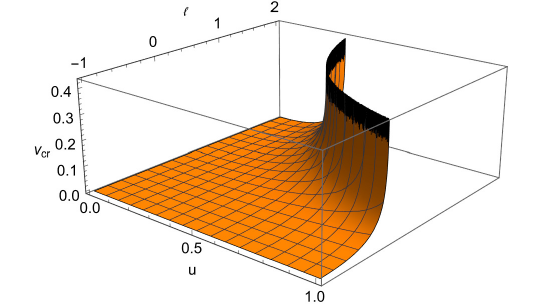}
\caption{The critical inclination angle parameter as the functions of $u$ and $\ell$.}
\label{fig10}
\end{figure}
By using Eqs.~(\ref{eq33}) and~(\ref{eq33-2}), the variation of the critical inclination angle parameter ${v_{\rm{cr}}}$ with respect to $\ell$ and $u$ is shown in Fig.~\ref{fig10}. It is evident that some values of $v$ and $u$ cannot be taken simultaneously. Therefore, to accurately analyze the general photon orbits of Kerr-like black hole in Einstein-Bumblebee gravity within the parameter space $\left(\ell, u \right)$, appropriate constraints on the parameters  $\ell$ and $u$ must be imposed to ensure physically meaningful parameter ranges and the absence of naked singularities. Specifically, the conditions $1 > v > 0$ and $1 > u > 0$ must be satisfied, and according to Eq.~(\ref{eq22}), to avoid the occurrence of naked singularities, the rotation parameter $\ell$ must satisfy the bound $\ell > -1$. Combining these conditions, the allowed parameter ranges are $-1 < \ell \leq 0 $, $ 0 \leq u \leq 1$  or $\ell > 0$, $0 \leq u \leq {1 \mathord{\left/ {\vphantom {1 {\left( {1 + \ell } \right)}}} \right.  \kern-\nulldelimiterspace} {\left( {1 + \ell } \right)}}$. These constraints ensure the physical validity of the photon orbit analysis and the consistency of the spacetime structure, providing a foundation for further investigations of photon orbits around the Kerr-like black hole in Einstein-Bumblebee gravity.

Now, substituting Eq.~(\ref{eq33-2}) into Eq.(\ref{eq33}) yields the critical inclination angle parameter. At this angle, the sextic~(\ref{eq23}) can be expressed as the product of a quartic $\mathcal{P}_4\left(x\right)$ and a quadratic $\mathcal{P}_2\left(x\right)$ as
\begin{align}
f\left(x\right) =
\mathcal{P}_4\left(x\right) \mathcal{P}_2\left(x\right) = \left(x^4 + A_1 x^3 + A_2 x^2 + A_3 x + A_4 \right)\left(x + A_5\right)^2,
\label{eq32}
\end{align}
where $A_1$, $A_2$, $A_3$, $A_4$, and $A_5$ are the coefficients consisting only of the rotation parameter $u$ and the LV parameter $\ell$.  By comparing with Eq.~(\ref{eq23}) with Eq.~(\ref{eq32}), the expressions of those coefficients are
\begin{subequations}
\label{eqa1}
\begin{align}
A_1 &= -4 - 2w, \\
A_2 &= \frac{3\left(2 + 8w + 3w^2 - 5w^3 + w^4\right)}{7 - 5w + w^2}, \\
A_3 &= \frac{6\left(2 - w - 2w^3 + w^4\right)}{7 - 5w + w^2}, \\
A_4 &= \frac{3\left(1 - w - w^3 + w^4 \right)}{7 - 5w + w^2}, \\
A_5 &= -1 + w.
\end{align}
\end{subequations}

\begin{figure}[htbp]
\centering
\includegraphics[width=0.6\linewidth]{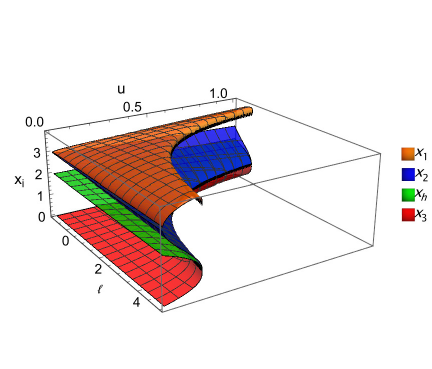}
\vspace{-1.5\baselineskip}
\caption{The general photon orbits $x_i$ and event horizon $x_h$ of the Kerr-like black hole in Einstein-bumblebee gravity as functions of the rotation parameter $u$ and LV parameter $\ell$.}
\label{fig11}
\end{figure}

By combining Eq.~(\ref{eqa1}) and Eq.~(\ref{eq23}), one can get the roots of the polynomial. Obviously, for the quartic polynomial part $\mathcal{P}_4\left(x\right)$, there exist three real roots, while the quadratic polynomial part $\mathcal{P}_2\left(x\right)$ has a double root denoted by $x = 1 - w$. Due to the length and complexity of these expressions, they are not explicitly presented here. Instead, numerical methods are used to illustrate the variation of these roots with respect to the parameters $u$ and $\ell$, as shown in Fig.~\ref{fig11}. It is found that the three real roots of the quartic part $\mathcal{P}_4\left(x\right)$, similar to the case of equatorial photon orbits, $x_1$ and $x_2$ lie outside the event horizon, while $x_3$ lies inside the event horizon $x_h$. The orange surface corresponding to $x_1$ is defined as the retrograde orbit since photons in this orbit move opposite to the rotation the black hole, requiring a larger orbital angular momentum, while the blue surface corresponding to $x_2$ represents the prograde orbit, where photons move in the same direction as the rotation the black hole. The green surface corresponds to $x_h$, and the red surface corresponds to $x_3$. Additionally, the double root of the quadratic part $\mathcal{P}_2\left(x\right)$ coincides with $x_3$.Similar to the cases of polar and equatorial orbits, when $\ell$ ranges from $-1$ to $0$, the parameter $u$ can take values from $0$ to $1$. However, when $\ell \geq 0$, the range of $u$ is constrained by $\ell$.

\begin{figure}[htbp]
\centering
\includegraphics[width=0.8\linewidth]{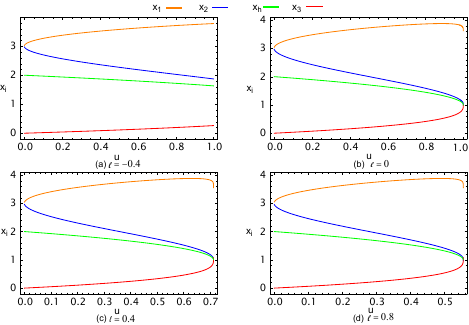}
\caption{The behavior of general photon orbits $x_i$ and event horizon $x_h$ versus the rotation parameter $u$ for different LV parameter $\ell$.}
\label{fig12}
\end{figure}

To facilitate the discussion of how the $\ell$ affects general photon orbits, we plot the behavior of $x_i$ as a function of $u$ for different $\ell$, derived from the roots of $\mathcal{P}_4\left(x\right)$ and $\mathcal{P}_2\left(x\right)$. In Fig.~\ref{fig12}(a), the rotation parameter $u$ is allowed to range from $0$ to $1$. The structure of photon orbits closely resembles that of the classical Kerr case (see Fig.~\ref{fig12}(b)). As $u$ increases, the retrograde orbit $x_1$ and the inner orbit $x_3$ increase monotonically, while the prograde orbit $x_2$ and the horizon $x_h$ decrease steadily. In Fig.~\ref{fig12}(c) and Fig.~\ref{fig12}(d), the allowed range of $u$ is restricted to $\ell  \le {1 \mathord{\left/ {\vphantom {1 {\left( {1 + \ell } \right)}}} \right. \kern-\nulldelimiterspace} {\left( {1 + \ell } \right)}}$. As $\ell$ increases, the photon orbits become less sensitive to changes in $u$. In particular, a subtle but significant phenomenon occurs near the extremal limit of $u$, the retrograde orbit $x_1$, instead of increasing monotonically, exhibits a peak and then decreases slightly as $u$ approaches its upper bound. This subtle behavior is clearly captured by the analytical method, while it is typically difficult to detect in numerical approaches used for black hole shadow studies. Additionally, the prograde photon orbit is located very close to the event horizon, making it possible to analyze the properties of the black hole's event horizon through observations of the prograde photon orbit.

\begin{figure}[htbp]
\centering
\includegraphics[width=0.6\linewidth]{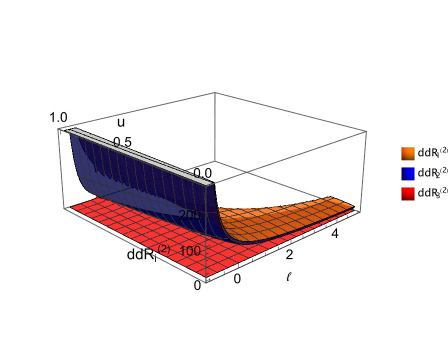}
\vspace{-1.5\baselineskip}
\caption{The ${\text{ddR}}_i^{\left( {2} \right)}$ as functions of $\ell$ and $u$ for general photon orbits $x_i$.}
\label{fig13}
\end{figure}

\begin{figure}[htbp]
\centering
\includegraphics[width=0.6\linewidth]{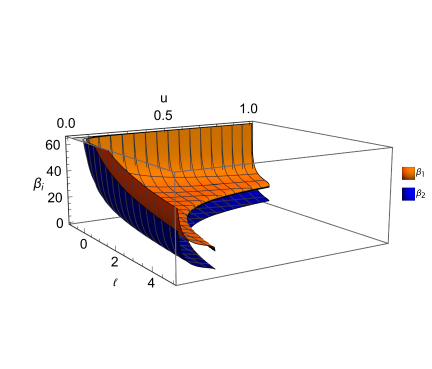}
\vspace{-1.5\baselineskip}
\caption{The critical impact parameters $\beta_i$  of photons on the general plane of the retrograde and prograde orbits as the functions of $\ell$ and $u$.}
\label{fig14}
\end{figure}

Finally, using Eq.~(\ref{eq30+}) and Eq.~(\ref{eq32}), we analyze the radial stability $\text{ddR}^{(2)}_i$ and the impact parameters $\beta_i$ for all general photon orbits. As shown in Fig.~\ref{fig13}, it is found that ${\text{ddR}}_1^{(2)} > 0$ and ${\text{ddR}}_2^{(2)} > 0$, while ${\text{ddR}}_3^{(2)} = 0$, indicating that the photon orbits $x_1$ and $x_2$ outside the event horizon are radially unstable, whereas the photon orbit $x_3$ inside the event horizon corresponds to a saddle point.  Fig.~\ref{fig14} shows that the impact parameters of the general photon orbits are similar to those of the equatorial plane. Therefore, the increasing of the rotation parameter $u$ and the LV parameter $\ell$ significantly reduces the retrograde orbit parameter $\beta_1$ and the prograde orbit parameter $\beta_2$, leading to considerable differences in the probability of photon scattering. As a result, the brightness of general photon orbits in Kerr-like  black hole in Einstein-bumblebee gravity is higher than that of the corresponding orbits of Kerr black hole. In addition, the parameter $\ell$ significantly restricts the range allowed by the rotation parameter $u$.


\section{Conclusion and discussion}
\label{sec5}
The study of photon orbits around a black hole is of great significance for understanding the structure of their gravitational fields and their physical properties. In this paper, we investigate the photon orbits around a Kerr-like black hole in Einstein-bumblebee gravity theory, focusing on the influence of the rotation parameter $u$, the LV parameter $\ell$, and the dimensionless parameter $v$ associated with the effective inclination angle on the orbit structure. By solving the Hamilton-Jacobi equation, we obtain a sixth-degree polynomial equation~(\ref{eq23}), and further classify photon orbits into three types: polar orbits, equatorial orbits, and general inclined orbits.

In the case of polar orbits ($v = 1$), two physically valid photon orbits are identified: $x_1$ is located outside the event horizon $x_h$, and $x_2$ is located inside it. As the LV parameter $\ell$ and the rotation parameter $u$ increase, the radii of $x_1$ and $x_h$ gradually decrease, while the radius of $x_2$ gradually increases. When $\ell$ ranges from $-1$ to $0$, the parameter $u$ can take values from $0$ to $1$. However, when $\ell \geq 0$, the allowed range of $u$ becomes increasingly constrained by $\ell$. This indicates that the parameter $\ell$ modifies the gravitational structure near the black hole, resulting in significant differences between the Kerr-like black hole and the classical Kerr black hole. In addition, it is found that the photon orbit $x_1$ outside the horizon is radially unstable, whereas the photon orbit $x_2$ inside the horizon corresponds to a saddle point. The critical impact parameter $\beta_i$ decreases with increasing $\ell$, implying that photons with smaller angular momentum are more easily captured by the black hole.

In the case of equatorial orbits ($v = 0$), three photon orbits exist: $x_1$ (retrograde) and $x_2$ (prograde), both located outside the event horizon, while $x_3$ lies inside the event horizon. As $\ell$ increases, the retrograde orbit $x_1$ and the inner orbit $x_3$ both increase in radius, while the prograde orbit $x_2$ gradually decreases. With increasing rotation parameter $u$, the radii of the event horizon and the prograde orbit decrease, whereas the radii of the retrograde orbit and the inner orbit $x_3$ increase. The photon orbits $x_1$ and $x_2$ outside the horizon are radially unstable, while the photon orbit $x_3$ inside the horizon corresponds to a saddle point. Similarly to the polar case, when $\ell$ ranges from $-1$ to $0$, the parameter $u$ can take values from $0$ to $1$, while for $\ell \geq 0$, the range of $u$ is restricted by $\ell$. Furthermore, the critical impact parameters of the photon orbits decrease with increasing $u$ or $\ell$, indicating that the increasing rotation and the LV parameter make photon capture less efficient. The critical impact parameter of the retrograde orbit is always greater than that of the prograde orbit, implying that the prograde orbit appears brighter than the retrograde one. These results suggest that equatorial photon orbits are highly sensitive to the parameter $\ell$, providing observable features to distinguish the Kerr-like black hole in nstein-Bumblebee gravity from the classical Kerr black hole.

In the general inclined case ($0 < v < 1$), in the extremal case, there exists a critical inclination angle $v_{\text{cr}}$ that governs the distribution of photon orbits. When $v > 3/7$ (i.e., $\cos i < \sqrt{4/7}$), regardless of the value of the rotation parameter $u$, two photon orbits are found outside the event horizon, namely the prograde orbit $x_1$ and the retrograde orbit $x_2$. This implies that at higher inclination angles, photons can orbit the black hole in both prograde and retrograde directions. In contrast, when $v < v_{\rm{cr}}$, only the orbit $x_1$ lies outside the horizon, while the orbit $x_2$ is either inside the horizon or coincides with it and therefore no longer exists as an independent stable orbit. It is evident that variations in the inclination angle significantly affect the structure and number of photon orbits, highlighting the dynamical corrections to photon motion introduced by the Bumblebee gravity theory. In the non-extremal case, within the parameter space $\left(v, u, \ell \right)$, there exist three photon orbits: the prograde orbit $x_1$ and the retrograde orbit $x_2$ are located outside the event horizon, while another orbit $x_3$ lies inside the horizon. Furthermore, similar to the polar and equatorial cases, the Lorentz-violating parameter $\ell$ imposes constraints on the allowable range of the rotation parameter $u$: when $\ell \leq 0$, $u$ can take any value in the range $0 < u < 1$, whereas for $\ell > 0$, the maximum value of $u$ is limited by $u \le {1 \mathord{\left/ {\vphantom {1 {\left( {1 + \ell } \right)}}} \right. \kern-\nulldelimiterspace} {\left( {1 + \ell } \right)}}$. Stability analysis shows that all photon orbits, whether polar, equatorial, or generally inclined, are radially unstable outside the event horizon. As $\ell$ increases, the critical impact parameter continuously decreases, indicating that photon orbits become increasingly sensitive to perturbations and deviations. This behavior implies that photon rings in Einstein-Bumblebee gravity may exhibit higher brightness and sharper structure than those in the classical Kerr case, offering potential signatures for future high-resolution astronomical observations.

\bibliography{references}

\end{document}